\documentstyle[aps,prb,psfig,floats]{revtex}
\begin{document}
%@twocolumnfalse\endcsname
\title{Longitudinal and transversal 
piezoresistive response of granular metals} 
\author{C. Grimaldi$^1$, P. Ryser$^1$, S. Str\"assler$^{1,2}$} 
\address{$^1$ D\'epartement de Microtechnique, IPM,
\'Ecole Polytechnique F\'ed\'erale de Lausanne,
CH-1015 Lausanne, Switzerland.}
\address{$^2$ Sensile Technologies SA, PSE, CH-1015 Lausanne, Switzerland.}

\maketitle

\centerline \\

\begin{abstract}
In this paper, we study the piezoresistive response and its anisotropy
for a bond percolation model of granular metals.
Both effective medium results and numerical Monte Carlo calculations
of finite simple cubic networks show that the piezoresistive anisotropy
is a strongly dependent function of bond probability $p$ and of
bond conductance distribution width $\Delta g$. 
We find that piezoresistive anisotropy
is strongly suppressed as $p$ is reduced and/or $\Delta g$ is enhanced
and that it vanishes at the percolation thresold $p=p_c$.
We argue that a measurement of the piezoresistive anisotropy could be
a sensitive tool to estimate critical metallic concentrations in 
real granular metals.

\end{abstract}
%\vskip 2pc ] 

%\narrowtext
\centerline \\

\section{introduction}
\label{intro} 

Granular metals are composite materials where a conducting phase
made of metallic particles or clusters with mean sizes ranging from
about 10 ${\rm \AA}$ to 200 ${\rm \AA}$ or more
is randomly dispersed into an insulating glassy material. 
Depending on the nature of the insulating
and conducting phases, their relative concentrations and the typical
size of the metallic granules, these materials show a quite rich
phenomenology and peculiar properties such as metal-insulator
transitions, giant Hall effect,\cite{pakhomov} and variable-range hopping
type of transport at low temperatures for insulating samples.\cite{mott}

The main ingredients governing most of the granular metal properties
are the concentration $x$ of metal in the composite, the activation
energy $E_c$ and the tunneling parameter $K=2d/\xi$, where $d$ is the
tunneling distance and $\xi$ is the localization length.
By varing $x$, the composite undergoes a metal-insulator transition
at a critical value $x_c$ which can be interpeted as the percolation
thresold.\cite{abeles} 
For $x>x_c$, connected metallic grains form a macroscopic cluster
which extends over the whole sample and current can flow easily from one end
to another when a voltage difference is applied. In this regime,transport
has a metallic character identified by a positive temperature
coefficient of resistivity (TCR). When $x<x_c$, transport is dominated by
tunneling and the resistivity increases as the temperature is lowered
(so TCR is negative). 
In this regime, the activation energy $E_c$ plays a fundamental role
since it defines the scale of energy an electron should overcome to
successfully tunnel from one grain to another. It is this parameter that
can give rise, together with $K$, to variable-range hopping type of 
transport at sufficiently low temperatures.\cite{mott,sheng}

The experimental determination of $x_c$ is however not as clear cut
as the above discussion seems to suggest. For granular metals with
low values of $E_c$ and $x$ close to the percolative thresold, TCR
can change sign as the temperature is varied in such a way that
${\rm TCR}<0$ for $T<T^*$ and ${\rm TCR}>0$ for $T>T^*$, where $T^*$ is some
temperature characteristic of the material.\cite{savvides} 
As pointed out in Ref.\onlinecite{chan},
the nonmonotonic behavior of TCR makes the determination of $x_c$
ambiguous and alternative methods should be developed.
 
In this paper we study the piezoresistive response of granular metals
and show how its anisotropy is strictly related to the vicinity to
the percolation critical point.
Piezoresistance, {\it i.e.}, the variation
of resistance $R$ or conductance $G$ upon an applied strain 
field $\varepsilon$,
is an effect common also in bulk metals. If we consider an isotropic
conducting material having bulk resistivity $\rho$, length $l$, and
cross-sectional area $A$, then the resistance $R$ is given by:
\begin{equation}
\label{R1}
R=\frac{\rho l}{A}.
\end{equation}
The fractional change of $R$, $\delta R/R$, due to a strain 
$\varepsilon=\delta l/l$ is given by:
\begin{equation}
\label{R2}
\frac{\delta R}{R}=\frac{\delta\rho}{\rho}+\frac{\delta l}{l}-
\frac{\delta A}{A}.
\end{equation}
If $\nu$ is the Poisson's ratio of the conductor, then it is possible
to show that $\delta A/A=-2\nu\varepsilon$ so that:
\begin{equation}
\label{R3}
{\rm GF}=\frac{\delta \rho}{\varepsilon\rho}+(1+2\nu),
\end{equation}
where we have defined the total piezoresistive gauge factor
${\rm GF}=\delta R/\varepsilon R$.\cite{white,prude1}
It is clear that the total piezoresistive response is made of an intrinsic,
$\delta\rho/\varepsilon\rho$, and a geometric, $1+2\nu$, component.
For bulk metals, GF is about 2 and, since $\nu$ is usually 
between 0.2 and 0.4, the geometrical effect is in this case
more important than the intrinsic one.\cite{white}

A different situation is encountered instead in the so-called
thick-film resistors, a particular
class of granular metals typically made of RuO$_2$, IrO$_2$, or 
Bi$_2$Ru$_2$O$_7$ granules  
embedded in a glassy matrix, for which GF values as large as $\sim 30$
has been reported,\cite{prude1} so that the intrinsic piezoresistive
effect is much larger than the geometrical one.
Such high strain-sensitivity resistances have been exploited to
manufacture piezoresistive sensor devices successfully used for
pressure and force measurements.\cite{white,prude1}
Due to their commercial applications, thick-film resistors are the granular
materials for which the piezoresistive effect has been studied best
and it is thought that inter-grain tunneling processes are responsible
for their GF values.\cite{pike,prude2} 
To illustrate this point, consider for example the
conductance $g$ of a simple tunneling process between two metallic grains: 
$g\sim \exp(-K)$. Under an applied strain, the tunneling distance $d$
is changed to $d+\delta d=d(1+\varepsilon)$ where $\varepsilon=\delta d/d$
is the strain. This modifies the tunneling parameter, $K\rightarrow
K+\delta K=K(1+\varepsilon)$ and consequently the conductance $g$.\cite{prude2}
The effect on $g$ can be substantial because of its exponential dependence
and, in principle, it can easily overwhelm the geometrical piezoresistive
effect. 

Of course, this tunneling mechanism of strain dependence should 
hold true also for other
granular metals than the thick-film resistors as long as tunneling
is an important element in transport properties. However the magnitude
of the intrinsic piezoresistive effect depends on several factors such as
elastic heterogeneity, diffusion of the metallic phase into the glass etc..

In this paper, we point out that a factor
affecting in an important way the piezoresistive response 
is the degree of tortuosity the current has in flowing through a sample
of granular metal. As an example, consider  Fig. 1 where we report the results
of a Monte Carlo calculation of a two dimensional bond-percolation
resistor-network model. In the network, a fraction $p$ of bonds distributed
at random has a finite conductance $g$, while the
remaining fraction $1-p$ has zero conductance. 
When all bonds of the network are conducting ($p=1$, Fig. 1a), 
the current flows along the direction of the
applied voltage difference, that is the $x$ direction. 
For $p<1$ the missing bonds
force the current to aquire nonzero components also along the direction 
perpendicular to the applied field (the $y$ direction) 
as shown in Fig. 1b where $p=0.8$.
For values of $p$ close to the percolation thresold $p_c$ 
($p_c=0.5$ for a simple square lattice), 
the tortuosity of the current flow is so developed that
the contributions of components parallel and perpendicular to the applied 
voltage drop become comparable. This is clearly shown in Fig. 1c where
the current flow is calculated for $p=0.55$.

How the tortuosity of the current flow affects the piezoresistive
response, or better its anisotropy, is qualitatively described as follows.
Immagine that each single bond in the direction $x$ of the applied field 
is slightly stretched so that its conductance becomes $g_x=g+g'_x\varepsilon$,
where $\varepsilon$ is the strain.\cite{note1} 
For simplicity, we keep the bonds along 
the direction perpendicular to the field unchanged: $g_y=g$.
The variation of the total conductance in this situation leads to the
longitudinal piezoresistive response (applied strain parallel to the
applied field).
Such longitudinal response is nonzero for all the three cases depicted 
in Fig. 1 since there are always nonzero
components of the current along the direction of the applied strain.
Consider now the transversal piezoresistive response for which the
strain is applied along the direction $y$ perpendicular to the field,
{\it i.e.}, $g_y=g+g'_y\varepsilon$ and $g_x=g$. Contrary to the
longitudinal case, for $p=1$ there are no components of the current
along the $y$ direction and the total conductance is therefore unaffected by
the applied strain. In this case the transersal piezoresistive 
response is zero. However, for $p<1$ current starts to flow also along the
$y$ direction and becomes affected by the strain induced variation of $g_y$
leading to a nonzero piezoristance effect. Since close to the percolation 
thresold the components of the current along the $x$ and $y$ directions
are comparable (Fig. 1c), it is expected that in this case
the longitudinal and the transversal piezoresistive responses have
similar values.

From the above qualitative discussion, we argue that away from the percolation
thresold the piezoresistive response is highly anisotropic with
the longitudinal component much larger than the transversal one,
while close to the percolative critical point the two components
becomes comparable, leading to an isotropic piezoresistive response.
By using Monte Carlo calculations and results from the effective medium theory
we show in the following that for resistor network models
this conclusion holds true and that actually
the longitudinal and the transversal piezoresistive responses
are equal at the percolation thresold. Hence, a measurement of the
piezoresistive anisotropy could be an alternative way to 
determine the critical concentration $x_c$ of the conducting phase
in granular metals. 

The paper is organized as follows. In the next section we introduce
the model and, by using the effective medium theory, provide analytical
formulas for the longitudinal and transversal piezoresistive responses.
In Sec. \ref{numerical} we report our Monte Carlo calculations and compare them
with the analytical formulas of the effective medium theory.
The last section is devoted to a discussion and to the conclusions.

\section{Effective medium theory}
\label{EMT}
In this section, we evaluate the piezoresistive response of a 
resistor network model by using a generalized effective medium theory.
Before entering the details of our model, let us first introduce
an operative definition of the piezoresistive effect and the different
gauge factors. In Fig. 2 we show a schematic apparatus for the
measurement of the piezoresistive response. The sample is placed on top of a
cantilever bar clamped at one end, and a force $F$ acts downward
on the opposite end. We assume that in the absence of the force, transport
is isotropic with conductance $G$. 
When $F\neq 0$, a tensile stress is built on the upper
surface of the cantilever and a strain $\varepsilon_{xx}=\varepsilon$
is transfered to the sample. This strain is directed along the main axis of 
the cantilever which we choose to be directed along the $x$ direction.
If the thickness of the cantilever is sufficiently small compared to its
width, the strain along $y$ can in first approximation be neglected.
If we assume that the strain field is completely transfered to the sample
(which is a good approximation as long as the linear dimensions of the
sample are much smaller than those of the cantilever),
then the strain field acting on the granular metal is:
$\varepsilon_{xx}=\varepsilon$, $\varepsilon_{yy}=0$, and
$\varepsilon_{zz}=-\varepsilon\nu/(1-\nu)$,
where $\nu$ is the Poisson ratio of the material. By referring to Fig. 2,
the longitudinal piezoresistive response is obtained by setting a
potential difference between points a and b, and the total conductance up 
to the linear term in the strain is:
\begin{equation}
\label{emt1}
G_x=G+\delta G_x=G(1-\varepsilon{\rm GF}_{\rm L}),
\end{equation}
where we have defined the longitudinal gauge factor:
\begin{equation}
\label{emt2}
{\rm GF}_{\rm L}=-\frac{\delta G_x}{\varepsilon G}.
\end{equation}
When the potential difference is applied between points c and d of
Fig. 2, that is perpendicular to the strain in the $x$ direction,
the conductance can be expressed as $G_y=G(1-\varepsilon{\rm GF}_{\rm T})$,
where ${\rm GF}_{\rm T}$ is the transversal gauge factor:
\begin{equation}
\label{emt3}
{\rm GF}_{\rm T}=-\frac{\delta G_y}{\varepsilon G}.
\end{equation}
Although ${\rm GF}_{\rm L}$ and ${\rm GF}_{\rm T}$ are the most used
parameters, there is also a vertical gauge factor obtained applying
a voltage difference between the upper and lower surfaces of the sample:
${\rm GF}_{\rm Z}=-\delta G_z/\varepsilon G$.

Let us describe now our resistor network model. We consider a three
dimensional simple cubic network where a bond has either
a finite conductance $g$ with probability $p$ or zero conductance
with probability $1-p$. We limit ourselves to the case in which the
activation energy $E_c$ is sufficiently low and the temperature
sufficiently high to approximate $g$ by a tunneling exponential.
Furthermore, we allow a certain distribution in the tunneling distance
in such a way that $g=g_0\exp[-K-\Delta K(p'-1/2)]$, where $g_0$
is a prefactor, $p'$ is a random variable uniformly distributed
between $0$ and $1$, and $\Delta K$ can assume values between zero
(single tunneling distance) and $\Delta K=2K$. For this latter case
therefore the tunneling exponent is continuously distributed between
$0$ and $K_{\rm max}=2K$. 

The probability distribution $\rho(g)$ of the
conductance value of a bond is:
\begin{equation}
\label{emt4}
\rho(g)=\frac{p}{\Delta K g}\,\Theta\!\left[g_0 e^{-K+\Delta K/2}-g\right]
\Theta\!\left[g-g_0 e^{-K-\Delta K/2}\right]
+(1-p)\delta(g),
\end{equation}
where $\Theta(x)$ is the Heaviside step function.
For a simple cubic network, the effective medium equation reads:\cite{kirk}
\begin{equation}
\label{emt5}
\int\! dg\, \rho(g)\frac{\bar{g}-g}{g+2\bar{g}}=0,
\end{equation}
which, by using Eq.(\ref{emt4}), leads to the following expression
for the effective macroscopic conductance $\bar{g}$:
\begin{equation}
\label{emt6}
\bar{g}=g_0\frac{e^{-K}}{2}\sinh\!\left( \frac{3p-1}{6p}
\Delta K\right)/\sinh\!\left(\frac{\Delta K}{6p}
\right).
\end{equation} 
As expected by the effective medium approximation, $\bar{g}$ 
vanishes for $p=\bar{p}_c=1/3$, which could be interpreted 
as the effective field metal-insulator transition point.

Now, Let us consider how the effective medium theory should be 
generalized to describe the effects of applied strain fields.
To simplify the problem, we make the assumption that the strains
are oriented along the directions of the bonds in the simple
cubic network. Hence, in the presence of the strain field, the
conductances become $g_x$, $g_y$, and $g_z$ according to whether
their corresponding conducting bonds are oriented towards to the $x$,
$y$, and $z$ directions, respectively. The probability distribution
$\rho_i(g)$ of a bond in the $i$ direction ($i=x,y,z$) is then
equal to Eq.(\ref{emt4}) where $K$ and $\Delta K$ are replaced
by $K_i$ and $\Delta K_i$, respectively.
Therefore, if $\varepsilon_{ii}=\delta d_i/d_i$
is the strain along the $i$ direction, $K_i=K(1+\varepsilon_{ii})$
and $\Delta K_i=\Delta K(1+\varepsilon_{ii})$. Within this model
we have made the assumption that the strain field does not
affect sensibly the prefactor $g_0$ and from now on we set it equal to unity.
The strain-induced anisotropy leads to a system of three coupled 
effective medium equations for the three macroscopic
conductances $\bar{g}_i$:\cite{berna}
\begin{equation}
\label{emt7}
\int\! dg\, \rho_i(g)\frac{\bar{g}_i-g}{g+S_i}=
\frac{\bar{g}_i}{S_i}+p\,\frac{\bar{g}_i+S_i}{S_i\Delta K_i}
\ln\!\left[\frac{S_i+e^{-(K_i+\Delta K_i/2)}}
{S_i+e^{-(K_i-\Delta K_i/2)}}\right]=0; 
\,\,\,\,\,\,\, i=x,y,z,
\end{equation}
where, by following the work of Bernasconi,\cite{berna} $S_i=1/R_i-\bar{g}_i$
and $R_i$ is the total resistance between two neighboring nodes in $i$ 
direction for the effective lattice. For the $x$ component, $R_x$ reads:
\begin{equation}
\label{emt8}
R_x=\frac{1}{\pi^3}\int_0^\pi\!\int_0^\pi\!\int_0^\pi 
dx'\,dy'\,dz'
\frac{1-\cos(x')}{\bar{g}_x[1-\cos(x')]+\bar{g}_y[1-\cos(y')]+
\bar{g}_z[1-\cos(z')]}, 
\end{equation}
and $R_y$ and $R_z$ are obtained from (\ref{emt8}) by replacing
in the numerator $\cos(x')$ with $\cos(y')$ and $\cos(z')$, respectively.
For general values of the $\bar{g}_i$, the integrals $R_i$ can be
evaluated only numerically.\cite{berna} 
However, we are interested in the 
piezoresistive effect which is just the linear response to an applied strain 
field so that, to our purposes, it is sufficient to evaluate the
integrals $R_i$ at the first order in the strain. As we show below
this procedure permits an analytical solution of the anisotropic
effective field equations.  

Let us replace in Eq.(\ref{emt8})
the effective conductances $\bar{g}_i$ with $\bar{g}_x=\bar{g}(1-
\varepsilon {\rm GF}_{\rm L})$, $\bar{g}_y=\bar{g}(1-
\varepsilon {\rm GF}_{\rm T})$, and  $\bar{g}_z=\bar{g}(1-
\varepsilon {\rm GF}_{\rm Z})$, where, 
by following the definitions given above, we have identified the gauge
factors for the effective conductances by:
\begin{equation}
\label{emt9}
{\rm GF}_{\rm L}=-\frac{\delta \bar{g}_x}
{\varepsilon \bar{g}};\,\,\,{\rm GF}_{\rm T}=
-\frac{\delta \bar{g}_y}
{\varepsilon \bar{g}};\,\,\,{\rm GF}_{\rm Z}=
-\frac{\delta \bar{g}_z}
{\varepsilon \bar{g}},
\end{equation}
As shown in the
Appendix, up to order $\varepsilon$, $R_x$ reduces to:
\begin{equation}
\label{emt10}
R_x=\frac{1}{3\bar{g}}+\frac{1-\xi}{3\bar{g}}{\rm GF}_{\rm L}
\varepsilon+\frac{\xi}{6\bar{g}}\left({\rm GF}_{\rm T}+
{\rm GF}_{\rm Z}\right)\varepsilon,
\end{equation}
where $\xi\simeq 0.5264$. Similarly, we obtain for $R_y$ and $R_z$:
\begin{eqnarray}
\label{emt11}
R_y &=& \frac{1}{3\bar{g}}+\frac{1-\xi}{3\bar{g}}{\rm GF}_{\rm T}
\varepsilon+\frac{\xi}{6\bar{g}}\left({\rm GF}_{\rm L}+
{\rm GF}_{\rm Z}\right)\varepsilon, \\
\label{emt11b}
R_z &=& \frac{1}{3\bar{g}}+\frac{1-\xi}{3\bar{g}}{\rm GF}_{\rm Z}
\varepsilon+\frac{\xi}{6\bar{g}}\left({\rm GF}_{\rm L}+
{\rm GF}_{\rm T}\right)\varepsilon.
\end{eqnarray}
By substituting Eqs.(\ref{emt10}-\ref{emt11b}) into Eq.(\ref{emt7}),
and equating to zero the term linear in $\varepsilon$, we obtain
three linear equations for the three gauge factors. After some algebra,
we obtain for the longitudinal and transversal components:
\begin{equation}
\label{emt12}
{\rm GF}_{\rm L}=\frac{[3p\Delta KX-6KpY-\Delta K]
[(1-2\nu)\xi (9pY-\Delta K)-12pY(1-\nu)]}
{18pY(1-\nu)[(4-9\xi)pY+\xi\Delta K]},
\end{equation}
\begin{equation}
\label{emt13}
{\rm GF}_{\rm T}=\frac{\xi[9pY-\Delta K]
[3p\Delta KX-6KpY-\Delta K](1-2\nu)}
{18pY(1-\nu)[(4-9\xi)pY+\xi\Delta K]},
\end{equation}
where:
\begin{eqnarray}
\label{emt14}
X&=&\frac{\sinh\left(\frac{\displaystyle \Delta K}{\displaystyle 6p}\right)
\cosh\left(\frac{\displaystyle 3p-1}{\displaystyle 6p}\Delta K\right)}
{\sinh\left(\frac{\displaystyle \Delta K}{\displaystyle 2}\right)} \\
\label{emt15}
Y&=&\frac{\sinh\left(\frac{\displaystyle \Delta K}{\displaystyle 6p}\right)
\sinh\left(\frac{\displaystyle 3p-1}{\displaystyle 6p}\Delta K\right)}
{\sinh\left(\frac{\displaystyle \Delta K}{\displaystyle 2}\right)}. 
\end{eqnarray}

We show in the next section how ${\rm GF}_{\rm L}$ and 
${\rm GF}_{\rm T}$ behave for different values of $p$, $K$,
and $\Delta K$ in comparison with Monte Carlo numerical calculations.
Before doing so, we stress here the main result of the effective medium theory,
namely, that ${\rm GF}_{\rm L}$ and ${\rm GF}_{\rm T}$
are generally different unless $p=\bar{p}_c=1/3$.\cite{note2} In fact, from
Eqs.(\ref{emt12}-\ref{emt15}) we obtain:
\begin{equation}
\label{emt16}
\lim_{p\rightarrow 1/3}{\rm GF}_{\rm L}=
\lim_{p\rightarrow 1/3}{\rm GF}_{\rm T}=
\frac{1-2\nu}{6(1-\nu)}\left[\Delta K\coth\!\left(\frac{\Delta K}{2}\right)
+2(K-1)\right].
\end{equation}
The effective medium theory therefore confirms the qualitative arguments
discussed in the introduction and, as we shall see in the next section,
it captures the essential physics governing piezoresistive anisotropy.
It is also worth stressing that, at $p=1/3$, ${\rm GF}_{\rm L}=
{\rm GF}_{\rm T}$ independently of the particular strain field used.

\section{Numerical calculations}
\label{numerical}

In this section we compare the results of the effective medium theory,
Eqs.(\ref{emt12}-\ref{emt15}), with numerical calculations of a three
dimensional random resistor network model.
In our Monte Carlo calculations, we have considered cubic networks 
with number of nodes up to $30\times 30\times 30$ and bond conductances
as described in the last section. The total conductance has been obtained
by solving numerically the Kirchoff equations for all nodes in the network,
and periodic boundary conditions have been imposed to the sides of the
network where no voltage difference is applied. The different gauge factors
have been obtained by calculating, for a fixed configuration of resistors,
the difference in conductances when $\varepsilon=0$ and
$\varepsilon=0.001$. The results have been then averaged over a minimum
of 30 to a maximum of 100 different runs.

In Fig. 3 we show the results of Monte Carlo calculations (symbols)
and effective medium theory (lines) for the limiting case for
which there is no distribution in tunneling probabilities ($\Delta K=0$).
In this limit, the effective medium formulas (\ref{emt12}-\ref{emt15})
simplify considerably: 
\begin{eqnarray}
\label{num1}
\lim_{\Delta K\rightarrow 0}{\rm GF}_{\rm L}&=&
K\frac{3\xi(1-p)(1-2\nu)-4(1-3p)(1-\nu)}
{(1-\nu)[9\xi(1-p)-4(1-3p)]}, \\
\label{num2}
\lim_{\Delta K\rightarrow 0}{\rm GF}_{\rm T}&=&
K\frac{3\xi(1-p)(1-2\nu)}
{(1-\nu)[9\xi(1-p)-4(1-3p)]}.
\end{eqnarray}
The results of Fig. 3 confirm the qualitative analysis made in the 
introduction (see also Fig. 1). Namely, for $p=1$ the current flows exclusively
along paths parallel to the direction $x$ of the applied field leading to
a non zero longitudinal piezoresistive effect (${\rm GF}_{\rm L}=K$)
and a vanishing transversal response (${\rm GF}_{\rm T}=0$). 
For $p<1$, the missing bonds force 
the current to flow also along directions perpendicular to $x$.
In this case, the piezoresistive response acquires a transversal component
while the longitudinal one is lowered. Note that the analytical and numerical
results agree quite well for bond probabilities larger than $p\sim 0.6$ while
for lower values of $p$ the effective medium results start to deviate
from the Monte Carlo data. This is of course due the increased inaccuracy of
the effective medium theory as the percolation thresold 
is approached. However, at the
critical bond probability ($p_c\simeq 0.25$ for simple cubic 
lattice,\cite{kirk} and $\bar{p}_c=1/3$ for the corrsponding
effective medium approximation)
both the Monte Carlo data and the analytical
results predict equal longitudinal and transversal responses.
Finally, note that for $\nu=0.5$, ${\rm GF}_{\rm T}$ remains equal to 
zero also for $p<1$. This is due to the fact that, for this value of the 
Poisson ratio and our assumption on the applied strain field,
one has $\varepsilon_{zz}=-\varepsilon$, therefore the bonds
in the $z$ direction are contracted of the same amount as the bonds in the
$x$ direction are stretched. 

The results of Fig. 3 show clearly how piezoresistive isotropy is
related to the vicinity to the percolation thresold. Another limiting 
situation is the one depicted in Fig. 4 where all bonds are present ($p=1$)
but with a finite ditribution of tunneling probability. Here, $\Delta K$
is varied between $\Delta K=0$ and $\Delta K=2K$, $K$ has been fixed
equal to $4$ and $\nu=0$, $0.3$, and $0.5$. 
For $\Delta K=0$ we recover the limit $p=1$ of Fig. 3,
while for finite values of $\Delta K$ the distribution of different tunneling
probabilities induces tortuosity in the current flow which is reflected
in a reduction of ${\rm GF}_{\rm L}$ and an enhancement of ${\rm GF}_{\rm T}$.
As before, the limit $\nu=0.5$ has the peculiarity of having
${\rm GF}_{\rm T}=0$ independently of the current tortuosity. Note that,
since for $K=4$ and $p=1$ the system is far away from the percolation thresold,
effective medium theory agrees excellently well with the Monte Carlo 
calculations. This agreement becomes however less satisfactory as the
distribution width increases as shown in Fig. 5 where results are reported 
for $\Delta K=0$, $K$, $2K$ as a function $K$ for $p=1$. High values of
tunneling distribution widths increase the fluctuactions of bond conductances
reducing the validity of the effective medium approximation.

Having analyzed two representative cases ($p\leq 1$ with $\Delta K=0$, 
and $p=1$ with $\Delta K \geq 0$) we show in Fig. 6 the results for both
$p\leq 1$ and $\Delta K\geq 0$ with $K=4$ and $\nu=0.3$.
The difference between the longitudinal and transversal piezoresistive 
effects decreases as the bond conductance fluctuations increases
whatever is the origin of such fluctuations (missing bonds or
finite distribution of bond conductances).
Note that, contrary to the other cases, for $\Delta K=8$ ${\rm GF}_{\rm L}$
increases as $p$ decreases. From the effective medium formulas it is
found that this feature persists also for higher $K$ values if $\Delta K$
is somewhat larger than $K$. However it is difficult to test this
tendency with our Monte Carlo calculations.since
we have employed the relaxation method for which 
large values of $\Delta K$ seriously slow the convergency of the 
recursive algorithm.

The effect of bond fluctuations on the piezoresistive anisotropy   
is more clearly shown in Fig. 7 where we plot, for the Monte Carlo 
data of Fig. 6, the quantity
$\chi$ defined as:
\begin{equation}
\label{chi}
\chi=\frac{{\rm GF}_{\rm L}-{\rm GF}_{\rm T}}{{\rm GF}_{\rm L}},
\end{equation}
which measures the degree of anisotropy of the piezoresistive response. 
For a fixed value of $\Delta K$, $\chi$ decreases as $p\rightarrow p_c$
from above, and it is expected to vanish at $p=p_c$. $\chi$
decreases also at fixed $p$ as $\Delta K$ is made larger.
We believe that $\chi$, because of its high-$p$ sensitivity for $p\sim p_c$,
could be a useful practical way to measure
the proximity of a granular metal to its percolation critical point.

\section{Discussion and conclusions}
\label{concl}

The results shown in the previous section clearly indicate
how the anisotropy of the piezoresistive effect depends on the
vicinity to the percolation thresold. We have interpreted the reduction
of anisotropy as due to the increasing of bond conductance fluctuations
as $p\rightarrow p_c$ from above. 
Although our results are quite general, their application to real 
granular metals needs some additional comments.

In our bond percolation model, we have assumed that the external 
strain fields are applied along the directions of the bonds in the network.
However, in a more realistic situation the bond directions should
be considered as random. In a practical calculation this could be 
achieved by considering an ensemble of ordered networks with different
orientations with respect to the applied strain fields. The average
over such an ensemble should be a satisfying description of a realistic case.
What we expect, and bond-average effective medium results (not reported here)
confirm, is that even for $\Delta K=0$ and $p=1$, the anisotropy 
parameter $\chi$, Eq.(\ref{chi}), is less than one. This is due to the random
bond orientation which contribute, even if the network have strictly equal
bond conductances, to a lowering of the piezoresistive anisotropy.
However, also in this case, the reduction of $\chi$ as $p$ is lowered (or
$\Delta K$ is enhanced) and the limit $\chi=0$ for $p\rightarrow p_c$
remain valid.

Another effect which has been ignored in the present analysis is the
possibility of having elastic heterogeneity within the granular metal.
A large difference between the elastic constants of the metallic and
insulating phases can give rise to important fluctuations in the
local strain fields. Since the microscopic tunneling processes are
affected by the local rather than the macroscopical strain values, 
the elastic heterogeneity can influence the piezoresistive response in 
an important way. This is actually what is expected in thick-film resistors
where very stiff metallic granules are embedded in a relatively elastically
soft insulating glass. This large elastic heterogeneity
could be at the origin of the large values of piezoresistive responses
actually observed for this class of granular materials.\cite{grima}
We expect however that the inclusion of this effect should influence
the absolute values of ${\rm GF}_{\rm L}$ and ${\rm GF}_{\rm T}$
but leave the anisotropy parameter $\chi$ relatively unaffected.

\begin{appendix}
\section{}
\label{app}

In this appendix we report the evaluation of the integrals $R_x$, $R_y$,
and $R_z$ up to the linear term in the strain $\varepsilon$. 
Let us consider first $R_x$, Eq.(\ref{emt8}), and substitute the effective
conductances $\bar{g}_i$, $i=x,y,z$, with  $\bar{g}_x=\bar{g}(1-
\varepsilon{\rm GF}_{\rm L})$, $\bar{g}_y=\bar{g}(1-
\varepsilon{\rm GF}_{\rm T})$, and  $\bar{g}_z=\bar{g}(1-
\varepsilon{\rm GF}_{\rm Z})$, where the gauge factors are
defined in Eq.(\ref{emt9}). Hence, up to the term linear in
$\varepsilon$ we obtain:
\begin{equation}
\label{a1}
R_x=\frac{a}{\bar{g}}+\frac{b}{\bar{g}}{\rm GF}_{\rm L}
\varepsilon+\frac{c}{\bar{g}}\left({\rm GF}_{\rm T}+
{\rm GF}_{\rm Z}\right)\varepsilon,
\end{equation}
where
\begin{eqnarray}
\label{a2}
a&=&\frac{1}{\pi^3}\int_0^\pi\int_0^\pi\int_0^\pi
dx'\,dy'\,dz'\,\frac{1-\cos(x')}{3-\cos(x')-\cos(y')-\cos(z')},\\
\label{a3}
b&=&\frac{1}{\pi^3}\int_0^\pi\int_0^\pi\int_0^\pi
dx'\,dy'\,dz'\,\frac{[1-\cos(x')]^2}{[3-\cos(x')-\cos(y')-\cos(z')]^2},\\
\label{a4}
c&=&\frac{1}{\pi^3}\int_0^\pi\int_0^\pi\int_0^\pi
dx'\,dy'\,dz'\,\frac{[1-\cos(x')][1-\cos(y')]]}
{[3-\cos(x')-\cos(y')-\cos(z')]^2}.
\end{eqnarray}
It is easy to show that $b+2c=a$ and $a=1/3$.
Moreover $b$ can be rewritten as $b=a-\xi/3$ where:
\begin{equation}
\label{a5}
\xi=\frac{1}{\pi^3}\int_0^\pi\int_0^\pi\int_0^\pi
dx'\,dy'\,dz'\,\frac{1}{3-\cos(x')-\cos(y')-\cos(z')}.
\end{equation}
Form Ref.\onlinecite{grad}, we have:
\begin{equation}
\label{a6}
\xi=\frac{4}{\pi^2}[18+12\sqrt{2}-10\sqrt{3}-7\sqrt{6}]
{\bf K}[(2-\sqrt{3})(\sqrt{3}-\sqrt{2})]^2 \simeq 0.5264,
\end{equation}
where ${\bf K}$ is the complete elliptic integral of the first kind.
Equation (\ref{emt10}) is obtained by
substituting $a=1/3$, $b=(1-\xi)/3$, and $c=\xi/6$ into Eq.(\ref{a1}).

\end{appendix}

\newpage

\begin{figure}
\protect
\centerline{\psfig{figure=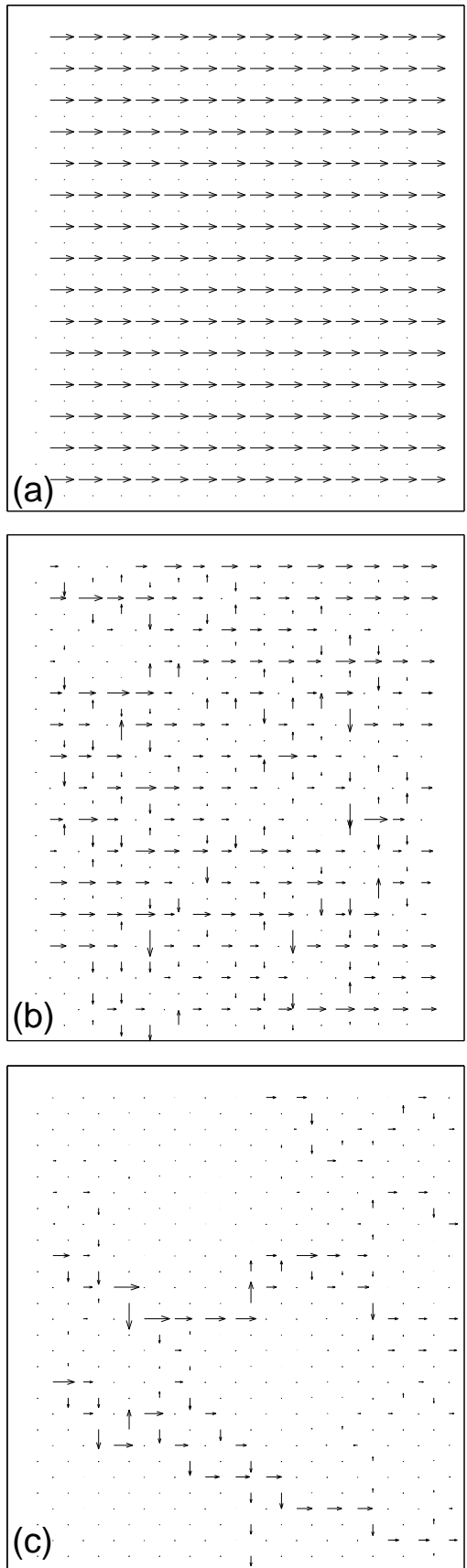,width=7cm}}
\caption{Calculated bond currents for a two-dimensional
bond percolation model. The external voltage difference
is applied between the left and right end sides of the 
network. The bond probabilities are $p=1$ (a),
$p=0.8$ (b), and $p=0.55$ (c). The directions and sizes
of the arrows represent the directions and the intensities
of the bond currents, respectively.}
\label{fig1}
\end{figure}

\begin{figure}
\protect
\centerline{\psfig{figure=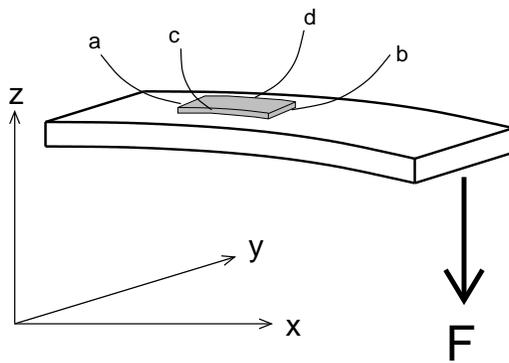,width=7cm}}
\caption{Schematical device for the measurement of the
piezoresistive response. The cantilever beam is clamped
at one end and a force $F$ acts downwards at the opposite
end. The resulting strain field is transfered to the 
material of interest (dark region). The longitudinal
response is obtained when a potential difference is
applied between points a and b and the resulting conductance
is measured for $F\neq 0$ and $F=0$. Instead, the transversal
response is obtained when the potential difference is applied
between c and d.}
\label{fig2}
\end{figure}

\begin{figure}
\protect
\centerline{\psfig{figure=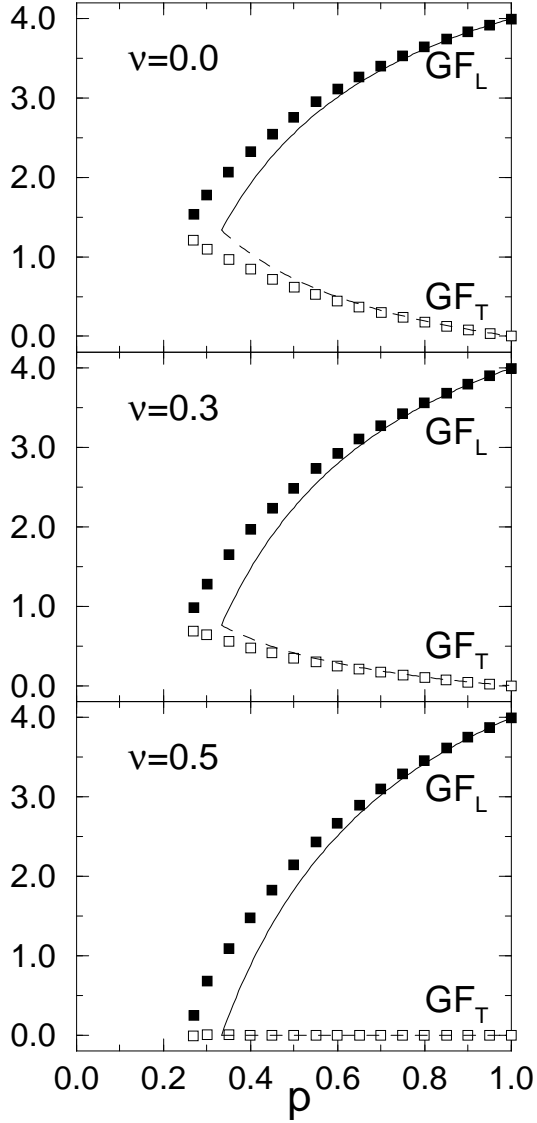,width=7cm}}
\caption{Longitudinal and tranversal gauge factors as a 
function of bond probability $p$ for different values of 
the Poisson ratio $\nu$. The tunneling probability has
zero distribution ($\Delta K=0$) and $K=4$.
Symbols refer to Monte carlo calculations (simple cubic
lattice of $30\times 30\times 30$ sites) while the solid
and dashed lines are the results of the effective medium 
approximation.}
\label{fig3}
\end{figure}

\begin{figure}
\protect
\centerline{\psfig{figure=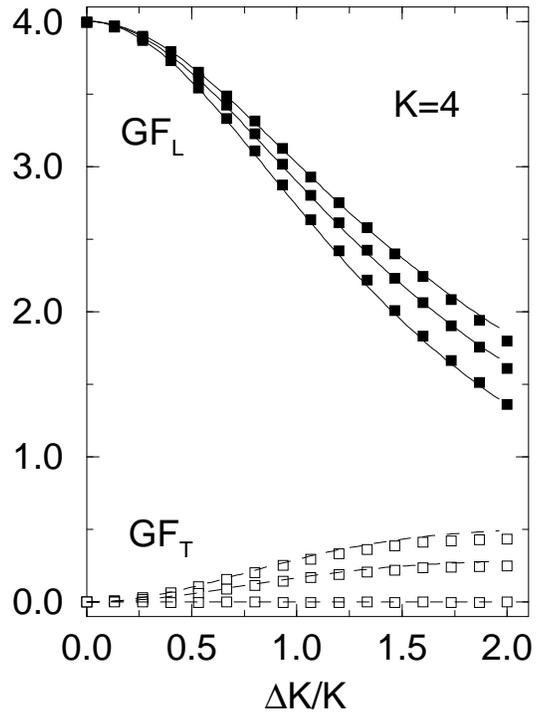,width=7cm}}
\caption{Longitudinal and tranversal gauge factors as a 
function of the tunneling distribution width $\Delta K$
for $K=4$ and $p=1$. From top to bottom: $\nu=0.0$, $0.3$,
and $0.5$. Symbols and lines have the same meaning as in Fig. 3
}
\label{fig4}
\end{figure}

\begin{figure}
\protect
\centerline{\psfig{figure=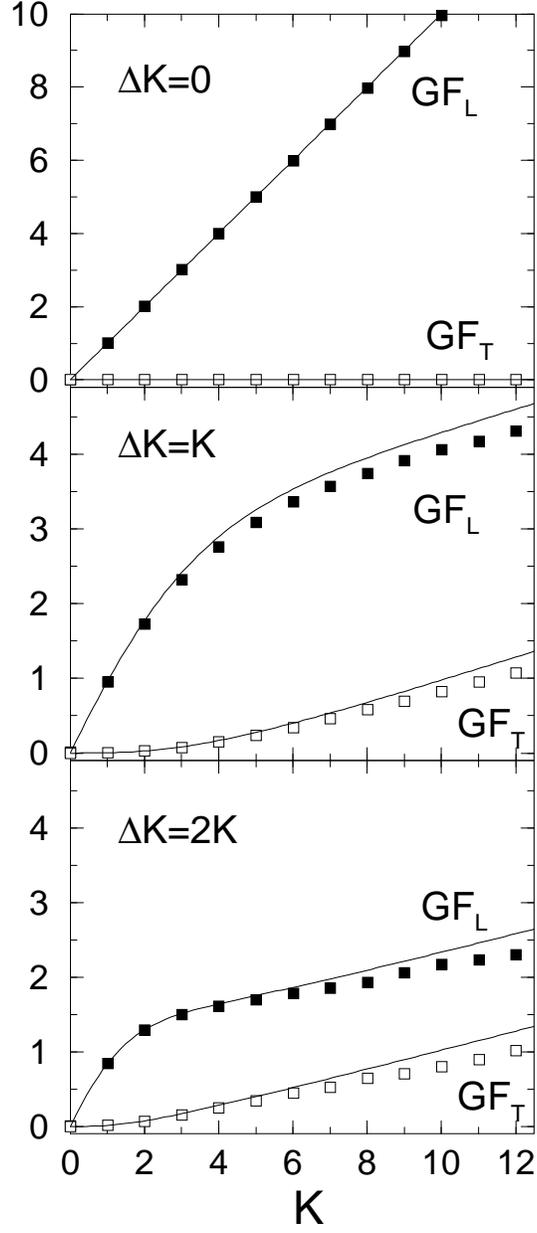,width=7cm}}
\caption{Longitudinal and tranversal gauge factors as a 
function of $K$ for $p=1$, $\nu=0.3$ and
different values of the tunneling distribution width $\Delta K$.
Symbols and lines have the same meaning as in Fig. 3
}
\label{fig5}
\end{figure}

\begin{figure}
\protect
\centerline{\psfig{figure=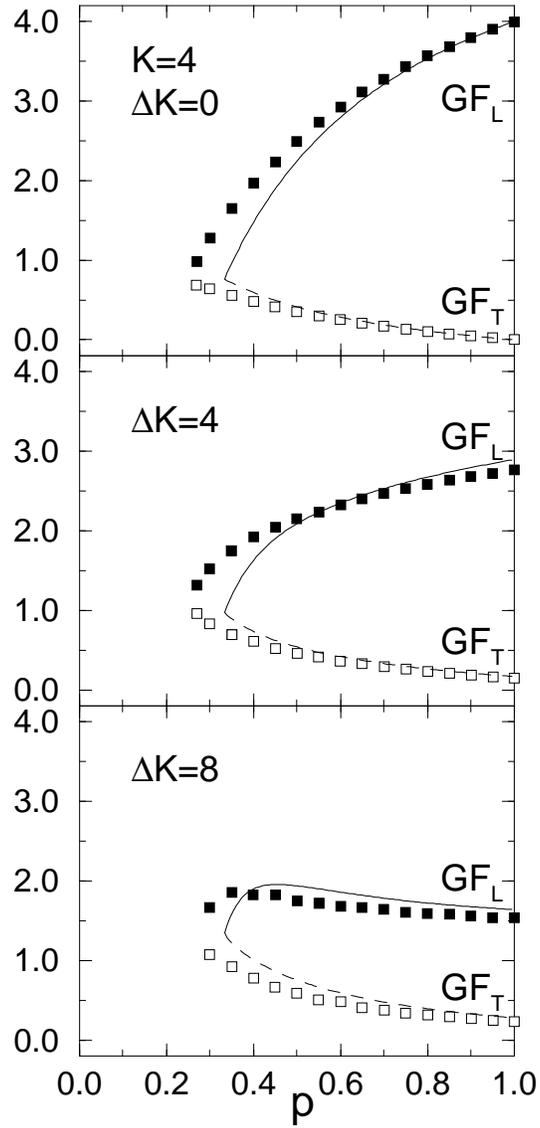,width=7cm}}
\caption{Longitudinal and tranversal gauge factors as a 
function of $p$ for
different values of the tunneling distribution width $\Delta K$.
$K=4$ and $\nu=0.3$.
Symbols and lines have the same meaning as in Fig. 3
}
\label{fig6}
\end{figure}

\begin{figure}
\protect
\centerline{\psfig{figure=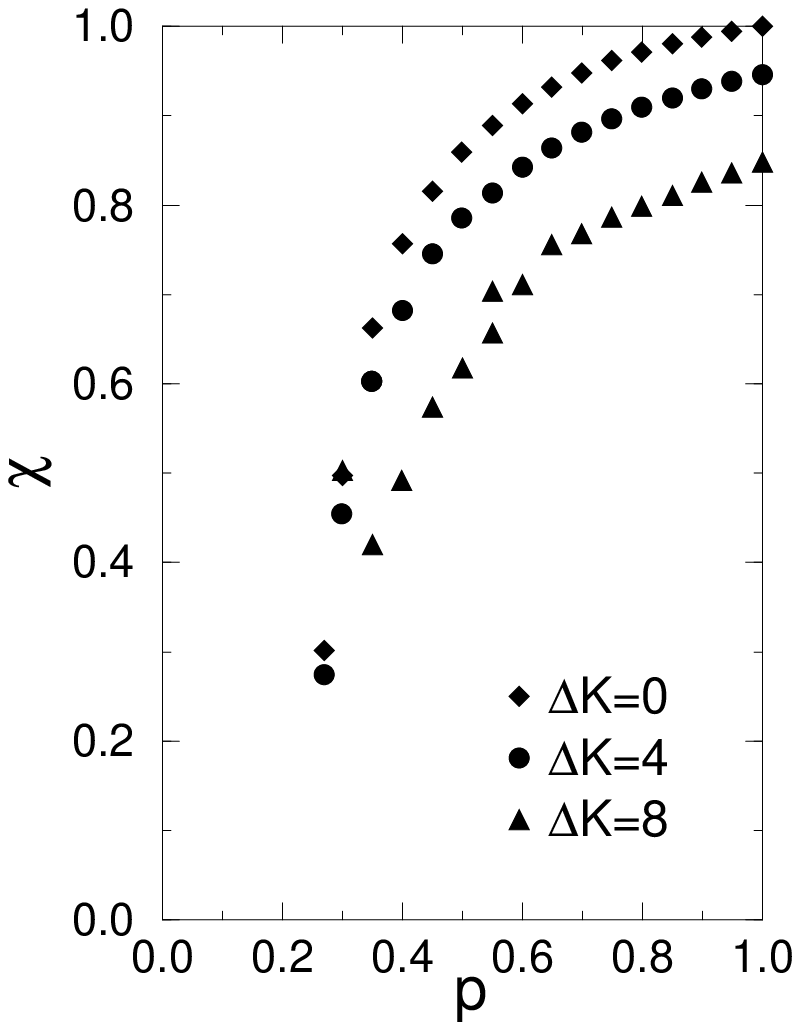,width=7cm}}
\caption{Factor of piezoresistive anisotropy $\chi$,
Eq.(\ref{chi}), for the Monte Carlo results of Fig. 6.}
\label{fig7}
\end{figure}


\begin{references}
\bibitem{note1}
The argument made in this qualitative discussion is general enough to leave
the specific strain dependence of the bond conductances unspecified.
\end{references}
\end{document}